\documentclass[11pt]{article}
\usepackage{amssymb,amsmath,cite,geometry}
\catcode`\@=11

\@addtoreset{equation}{section}
\def\theequation{\arabic{section}.\arabic{equation}}
     
\catcode`\@=11
\def\thesection{\arabic{section}}

\def\appendix{\setcounter{section}{0}
        \def\thesection{Appendix.}
        \def\theequation{\Alph{section}.\arabic{equation}}}
\def\section{\@startsection{section}{1}{\z@}{3.5ex plus 1ex minus
  .2ex}{2.3ex plus .2ex}{\large\bf}}
\makeatletter
\renewcommand{\@seccntformat}[1]{\csname the#1\endcsname.\quad}
\makeatother

\long\def\@makefntext#1{\parindent 0cm\noindent
\hbox to 1em{\hss$^{\@thefnmark}$}#1}

\newcommand{\captionfonts}{\small}
\makeatletter  % Allow the use of @ in command names
\long\def\@makecaption#1#2{%
  \vskip\abovecaptionskip
  \sbox\@tempboxa{{\captionfonts #1: #2}}%
  \ifdim \wd\@tempboxa >\hsize
    {\captionfonts #1: #2\par}
  \else
    \hbox to\hsize{\hfil\box\@tempboxa\hfil}%
  \fi
  \vskip\belowcaptionskip}
\makeatother   % Cancel the effect of \makeatletter 

\begin{document}
\begin{titlepage}
\vspace{.5in}
\begin{flushright}
October 2014\\  %date
\end{flushright}
\vspace{.5in}
\begin{center}
{\Large\bf
 A Note on Black Hole Entropy\\[1ex] in Loop Quantum Gravity}\\  %title
\vspace{.4in}
{S.~C{\sc arlip}\footnote{\it email: carlip@physics.ucdavis.edu}\\
       {\small\it Department of Physics}\\
       {\small\it University of California}\\
       {\small\it Davis, CA 95616}\\{\small\it USA}}
\end{center}

\vspace{.5in}
\begin{center}
{\large\bf Abstract}
\end{center}
\begin{center}
\begin{minipage}{4.95in}
{\small
Several recent results have hinted that black hole 
thermodynamics in loop quantum gravity simplifies if one chooses an
imaginary Barbero-Immirzi parameter $\gamma=i$.  This suggests a
connection with $\mathrm{SL}(2,\mathbb{C})$ or $\mathrm{SL}(2,\mathbb{R})$
conformal field theories at the ``boundaries'' formed by spin network
edges intersecting the horizon.  I present a bit of background regarding the
relevant conformal field theories, along with some speculations about how 
they might be used to count black hole states.  I show, in particular,
that a set of unproven but plausible assumptions can lead to a 
boundary conformal field theory whose density of states matches the
Bekenstein-Hawking entropy.
}
\end{minipage}
\end{center}
\end{titlepage}
\addtocounter{footnote}{-1}

The computation of black hole entropy \cite{Rovelli,ABCK,ABK} has been one
of the triumphs of loop quantum gravity.  The achievement has come at a
price, however: the area operator in loop quantum gravity depends on
the Barbero-Immirzi parameter $\gamma$, and one must tune $\gamma$
to a rather peculiar value, fixed by an obscure combinatorial problem, 
to obtain the correct Bekenstein-Hawking entropy \cite{Domagala,Meissner}.
This tuning only has to be done once---if $\gamma$ is adjusted to
fit, say, the Schwarzschild black hole, the theory gives the correct entropy
for all black holes---but the physical meaning of this choice remains
enigmatic.

Over the past few years, some interesting new hints have emerged.  
First, it has been observed that in a grand canonical ensemble, the dependence 
of entropy on $\gamma$ can be shifted to a chemical potential term 
for the number of ``punctures'' where a spin network intersects the horizon 
\cite{Ghosh}.  In the presence of suitable holographic matter fields, it
may be possible to set this chemical potential to zero 
\cite{Perez}, and steps have been taken to develop a conformal field 
theory description \cite{GhoshPran}.  At the same time, it has been noted 
that if one analytically continues the Barbero-Immirzi parameter to the value
 $\gamma=i$, certain partition functions automatically give the correct 
Bekenstein-Hawking entropy \cite{Geiller,Achour}, and certain spin
foam amplitudes acquire a correct imaginary part \cite{Bodendorfer}.
The choice $\gamma = \pm i$ is the one for which Ashtekar variables 
were first introduced \cite{Ashtekar}, and it is a natural one: the 
loop quantum gravity connection is self-dual at this value, and is the 
only value for which the Barbero-Ashtekar
connection is a diffeomorphism-invariant spacetime gauge field 
\cite{Samuel,Alexa}.  Unfortunately, it is not clear how to consistently
implement reality conditions when $\gamma$ is complex; at
least for practical purposes, a real value seems to be needed.

I do not have the answers to these problems, and it remains possible
that a correct thermodynamic treatment could give a value of the
entropy that is independent of $\gamma$ \cite{Bianchi}.  But in this 
note I will suggest some ingredients that might be included in
the answers, including a few elements of conformal field theory that
may not be familiar to most people working in loop quantum gravity.
In particular, there are interesting hints that coset constructions, 
Liouville theory, and holomorphic Chern-Simons theories may be 
relevant.

\section{Chern-Simons and WZNW theory \label{CSWZW}}

A question about a black hole in quantum gravity is a question about
conditional probability: one must impose the condition that the
desired black hole is present.  One way to do so is to demand that
a hypersurface $\Delta$ be a horizon.  Such a horizon is not a 
physical boundary, but it is a place where one is imposing
``boundary conditions,'' and as a consequence, the Einstein-Hilbert 
action acquires an added boundary term.  In the first-order 
formalism of loop quantum gravity, this term is a three-dimensional 
Chern-Simons action \cite{ABCK,ISO,Krasnov0},
\begin{align}
I_{\mathit{CS}} = \frac{k}{4\pi} \int_\Delta \mathrm{Tr}
  \left\{A\wedge dA + \frac{2}{3} A\wedge A\wedge A \right\} ,
\label{a1}
\end{align}
where $A = A_\mu{}^a T_a\,dx^\mu$ is a Lie-algebra-valued connection
one-form.\footnote{The appearance of a Chern-Simons boundary 
action was first suggested in a slightly different context by Smolin 
\cite{Smolin0,Smolinb}.  An related $\mathrm{SL}(2,\mathbb{C})$
Chern-Simons action also appears in certain spin foam amplitudes
for (3+1)-dimensional gravity with a positive cosmological constant
\cite{Haggard}, although the connection to black hole horizons is not
yet clear.}  In loop quantum gravity, it is usual to start with a 
self-dual $\mathrm{SL}(2,\mathbb{C})$ connection, making (\ref{a1}) a 
holomorphic $\mathrm{SL}(2,\mathbb{C})$ Chern-Simons action.
(The general $\mathrm{SL}(2,\mathbb{C})$ Chern-Simons action 
contains a second term involving the complex conjugate $\bar A$.)
From the point of view of Chern-Simons theory, this choice---and
thus the choice $\gamma=i$---is special: holomorphic Chern-Simons 
actions are expected to have exceptional properties, but they are also 
exceptionally poorly understood \cite{Wittend}.

Chern-Simons theory is intimately related to a two-dimensional
conformal field theory, Wess-Zumino-Novikov-Witten (WZNW, or WZW) 
theory \cite{Wittena}.  To see this, consider the case that $\Delta$ itself 
has a boundary, which may again be merely a ``place  where one
imposes boundary conditions.''   The action (\ref{a1}) must now
itself acquire a boundary term: under a variation of $A$, we have
$$
\delta I_{\mathit{CS}} 
  = \frac{k}{2\pi}\int_\Delta \mathrm{Tr}\left[\delta A\left(dA + A\wedge A\right)\right]
            -\frac{k}{4\pi}\int_{\partial\Delta} \mathrm{Tr}\left(A\wedge\delta A\right) ,
$$
and we must modify the action to cancel the boundary variation.  
To simplify comparison with the literature, choose complex coordinates 
$\{z,{\bar z}\}$ on $\partial\Delta$, and fix $A_z$ and the boundary.\footnote{This
choice simplifies notation, but can be easily generalized.  For each value of
the Lie algebra index $a$, $A^a_z$ and $A^a_{\bar z}$ are canonically
conjugate; one can fix any linear combination of these at the boundary and 
leave the conjugate component free.}  The required boundary term is then
\begin{align}
I_{\mathit{bdry}}[A] = \frac{k}{4\pi}\int_{\partial\Delta}\mathrm{Tr} A_zA_{\bar z} .
\label{a2}
\end{align}

We now make the crucial observation that the total action (\ref{a1}--\ref{a2})
is not gauge invariant at the boundary \cite{Ogura,Carlipi}.   If we write $A$ as
a gauge-fixed value $\bar A$ and a gauge transformation $g(z,{\bar z})$,
\begin{align}
A = g^{-1}dg + g^{-1}{\bar A}g ,
\label{a3}
\end{align}
we find that
\begin{align}
(I_{\mathit{CS}}+I_{\mathit{bdry}})[A] = (I_{\mathit{CS}}+I_{\mathit{bdry}})[{\bar A}]
  + k I^+_{\mathit{WZW}}[g^{-1},{\bar A}] ,
\label{a4}
\end{align}
with
\begin{align}
I^+ _{\mathit{WZNW}}[g^{-1},{\bar A}_z]
 = \frac{1}{4\pi}\int_{\partial\Delta}\mathrm{Tr}
 \left(g^{-1}\partial_z g\,g^{-1}\partial_{\bar z}g
 - 2g^{-1}\partial_{\bar z}g {\bar A}_z\right)
 + \frac{1}{12\pi}\int_\Delta\mathrm{Tr}\left(g^{-1}dg\right)^3   .
\label{a5}
\end{align}
This is precisely chiral WZNW action for a field $g$ coupled to a background 
gauge potential ${\bar A}_z$.  

The gauge transformation $g$ has thus become dynamical at the
boundary.  In a reasonable sense, in fact, it comprises most of the 
degrees of freedom: a Chern-Simons theory on a
 closed manifold is a topological field theory with only finitely many degrees 
of freedom, while a WZNW model is a true field theory, albeit in one dimension
 less.  The boundary degrees of freedom have a physical interpretation as 
``would-be gauge degrees of freedom'' \cite{Carlipa}, degrees of freedom 
that are only physical because the boundary conditions break 
the gauge invariance of the bulk.  Their presence is necessary for
consistency---for instance, if one ``glues'' two manifolds
along a common boundary, the path integrals for Chern-Simons 
theories on each half combine correctly only if one includes the
boundary WZNW model \cite{Wittenb}.

For black holes, two types of boundaries of $\Delta$ are of particular
interest.  Suppose that $\Delta$ initially has the usual horizon topology
$\mathbb{R}\times S^2$.  If we cut a ``hole'' in a spherical 
cross-section of $\Delta$, the resulting boundary will be a cylinder
$\partial\Delta\approx \mathbb{R}\times S^1$.  The dynamical variable 
in (\ref{a5}) will be a group-valued field $g(\phi,v)$, and (\ref{a5}) will 
describe a WZNW model on a cylinder.  This is a well-understood quantum
theory \cite{Wittenf}, which can be quantized, for instance, in terms of 
loop groups \cite{EMSS}.  

Suppose, on the other hand, that we shrink the ``hole'' to a point $p$,
a ``puncture.''  The boundary $\mathbb{R}\times\{p\}$ is now a line, 
which can be interpreted in the Chern-Simons theory as a Wilson line.
The field $g$ no longer has any spatial dependence; the only remaining
information is the holonomy around the one-dimensional boundary,
and the quantum theory shrinks to a theory of the conformal blocks
of the WZNW model \cite{Wittena,EMSS}.

\section{Virasoro algebras and central charge \label{VA}}

The WZNW model (\ref{a5}) is a two-dimensional conformal field
theory.   Such theories have a very powerful infinite-dimensional 
symmetry group, which determines much of their behavior.
The metric for a two-dimensional manifold can always be written 
locally as
$$
ds^2 = 2g_{z{\bar z}}dzd{\bar z}
$$
where the complex coordinates $\{z,{\bar z}\}$ were introduced  
in the preceding section.  The
holomorphic and antiholomorphic diffeomorphisms $z\rightarrow z + \xi(z)$,
${\bar z}\rightarrow {\bar z} + {\bar\xi}({\bar z})$ rescale the 
metric, and a conformal field theory is invariant under such rescalings.   
The canonical generators of this symmetry, denoted $L[\xi]$ and 
${\bar L}[{\bar\xi}]$, satisfy a Virasoro algebra \cite{FMS},
\begin{subequations}
\begin{align}
&[L[\xi],L[\eta]] 
   = L[\eta\xi' - \xi\eta']   + \frac{c}{48\pi}\int dz\left( \eta'\xi'' - \xi'\eta''\right)
  \label{b1a}\\
&[{\bar L}[{\bar\xi}],{\bar L}[{\bar\eta}]] 
   = {\bar L}[{\bar\eta}{\bar\xi}' - {\bar\xi}{\bar\eta}']  + \frac{{\bar c}}{48\pi}%
   \int d{\bar z}\left( {\bar\eta}'{\bar\xi}'' - {\bar\xi}'{\bar\eta}''\right) 
\label{b1b}\\
&[ L[\xi],{\bar L}[{\bar\eta}]] = 0 . \label{b1c}
\end{align}
\end{subequations}
The first terms on the right-hand sides of (\ref{b1a}--\ref{b1b}) give the usual algebra 
of diffeomorphisms.  The remaining terms provide a unique central extension,
fixed by the values of the two constants $c$ and $\bar c$, the 
central charges.  As usual, the zero modes $\Delta_0$ and ${\bar\Delta}_0$
of $L_0$ and ${\bar L}_0$ are conserved quantities, the ``conformal weights,'' 
which can be viewed as linear combinations of mass and angular momentum.

Remarkably, Cardy has shown that with a few mild restrictions, the asymptotic 
density of states of any two-dimensional conformal field theory is almost
completely determined by the symmetries \cite{Cardy,Cardy2}.  Let 
$\Delta_{\hbox{\scriptsize\it min}}$ and ${\bar\Delta}_{\hbox{\scriptsize\it min}}$
be the lowest eigenvalues of $L_0$ and ${\bar L}_0$ (usually but not always
zero), and define
\begin{align}
c_{\hbox{\scriptsize\it eff}} = c - 24\Delta_{\hbox{\scriptsize\it min}}, \qquad
{\bar c}_{\hbox{\scriptsize\it eff}} = {\bar c} - 24{\bar\Delta}_{\hbox{\scriptsize\it min}} .
\label{b2}
\end{align}
Then the density of states $\rho$ in a microcanonical ensemble, at fixed eigenvalues
$\Delta$ and $\bar\Delta$ of $L_0$ and ${\bar L}_0$, behaves as \cite{Carlipc}
\begin{align}
\ln\rho(\Delta) 
  \sim 2\pi\sqrt{\frac{c_{\hbox{\scriptsize\it eff}}\Delta}{6}}, \qquad 
\ln{\bar\rho}({\bar\Delta}) 
  \sim 2\pi\sqrt{\frac{{\bar c}_{\hbox{\scriptsize\it eff}}{\bar\Delta}}{6}}  
\label{b3}
\end{align}
and in a canonical ensemble, at fixed temperature $T$, as \cite{BoussoM}
\begin{align}
\ln\rho(T) \sim \frac{\pi^2}{3}c_{\hbox{\scriptsize\it eff}}\,T, \qquad 
\ln{\bar\rho}(T) \sim \frac{\pi^2}{3}{\bar c}_{\hbox{\scriptsize\it eff}}\,T 
\label{b4}
\end{align}
where $T$ is 
a dimensionless temperature, determined by the periodicity of a dimensionless
conformal time.  (Under some circumstances, the ``left temperature'' $T$ 
and ``right temperature'' $\bar T$ may also differ.)  The entropy is thus 
determined by a few parameters, independent of the details of the theory.

There is one remaining subtlety:
a two-dimensional conformal field theory may be defined on the complex 
plane or on a cylinder.  The two are related by the transformation $w = \ln z$,
but because of the conformal anomaly, the Virasoro generators do not quite
transform as tensors \cite{FMS}. Rather,
\begin{align}
\Delta_{\hbox{\scriptsize\it cyl}} 
    = \Delta_{\hbox{\scriptsize\it plane}} - \frac{c}{24}, \qquad
{\bar\Delta}_{\hbox{\scriptsize\it cyl}} 
    = {\bar\Delta}_{\hbox{\scriptsize\it plane}} - \frac{{\bar c}}{24}  .
\label{b5}
\end{align}
The conformal weights $\Delta$ and $\bar\Delta$ appearing in 
 (\ref{b3})--(\ref{b4}) are those for the cylinder, as may be checked from 
the derivation in \cite{Carlipc}.

\section{$\mathbf{SL}(2)$ connections, coset constructions,
and Liouville theory \label{SL2}}

In the first order formulation of general relativity, in addition to the
diffeomorphisms, the fundamental symmetries include local Lorentz 
transformations, which take their values in $\mathrm{SL}(2,\mathbb{C})$.  
In the usual approach to loop quantum gravity, one gauge-fixes the latter 
to $\mathrm{SU}(2)$, the little group of the timelike normal to a spatial 
slice.   The boundary Chern-Simons theory is then an $\mathrm{SU}(2)$
theory---perhaps further reducible to $\mathrm{U}(1)$ \cite{ABK}%
---with a real Barbero-Immirzi parameter.  Both the area of the
horizon and the dimension of the Chern-Simons Hilbert space are
determined by the  spins carried by spin network edges crossing
the horizon.  Determining the relationship between the two---%
and thus the Bekenstein-Hawking  entropy---becomes a combinatorial
problem; see, e.g., \cite{ENPP} for a careful treatment in the 
$\mathrm{SU}(2)$ case.

As noted above, though, this choice requires an awkward fine-tuning 
of the Barbero-Immirzi parameter, and there are hints that the picture
may simplify if one can instead choose  $\gamma=i$.  This requires
a complex connection, and suggests that we should look at 
the full gauge group of self-dual connections, complexified 
$\mathrm{SU}(2)$,  which is isomorphic to $\mathrm{SL}(2,\mathbb{C})$.  
This is the group that seems most natural in light of the linear
simplicity constraints of spin foam theory \cite{EPRL}
\begin{align}
K^i = \gamma L^i
\label{c0}
\end{align}
with $\gamma=i$, where the $K^i$ and $L^i$ are generators of 
``internal'' boosts and rotations. 

$\mathrm{SL}(2,\mathbb{C})$ may be too large a group, however.  
The analytic continuation to imaginary $\gamma$ in \cite{Geiller,Achour} 
is really a continuation to an $\mathrm{SL}(2,\mathbb{R})$ (or, 
equivalently, $\mathrm{SU}(1,1)$) algebra.  Frodden et al.\ \cite{Geiller} 
argue that this subgroup is picked out by a reality condition for the area; 
it is also the little group for a unit normal to a stretched horizon.\footnote{%
There are also arguments for a smaller $\mathrm{ISO}(2)$ or 
$\mathrm{ISO}(1,1)$ gauge group \cite{Kaul,Wang}.}

For now, both of these possibilities seem worth considering.  Of course, 
this is not trivial---in neither case do we know how to quantize the 
theory, except in some sense as an analytic continuation from real 
$\gamma$.  Still, it is worth seeing where these alternatives might take
us.   We might at first expect that such choices should be irrelevant: 
unless there is an anomaly, it should not matter how we 
gauge-fix a symmetry.  But as noted earlier, boundary conditions
can break the symmetry, elevating ``would-be gauge
transformations'' to true degrees of freedom.   If these are indeed
relevant degrees of freedom, the choice of
gauge group and the pattern of symmetry-breaking  
could be crucial.

From section \ref{CSWZW}, we now expect an $\mathrm{SL}(2,\mathbb{C})$
or $\mathrm{SL}(2,\mathbb{R})$ Chern-Simons theory on the horizon.  
Neither of these is a very well-understood theory (though see \cite{Wittend,%
Wittene,MalOog,MalOogb}).  Fortunately, though, there is a potential
simplification: we have not yet exhausted our use of boundary conditions.

An isolated horizon \cite{ISO} is characterized by a null normal $\ell^a$ 
satisfying the condition $\ell^b\nabla_b\ell^a = \kappa\ell^a$ on $\Delta$.
Following \cite{IH}, let us choose a constant ``internal'' vector $\ell^I$
such that $\ell^a = e^a{}_I\ell^I$.  The spin connection $\omega$
must then satisfy $\ell^a\omega_a{}^I{}_J\ell^J = \kappa\ell^I$ at the horizon, 
and it may be checked that this implies 
\begin{align}
A_v{}^I{}_J\ell^J \propto \ell^I
\label{c2}
\end{align}
for the full self-dual connection.
From (\ref{a3}), this means that certain components of the WZNW current
$$
(J_v)^I{}_+ = k\left(g^{-1}\partial_vg + g^{-1}{\bar A}_vg\right)^I{}_+
$$
are fixed.  

Modifications of this type---in which a current corresponding to a 
subgroup of the gauge group is held fixed---have been the subject
of extensive study.   The resulting conformal field theories are known 
as coset models \cite{FMS,Gawedzki}.  In our case, the relevant coset is 
$\mathrm{SL}(2,\mathbb{C})/E(2)$, where $E(2)$ is the little group of
the null vector $\ell^A$, the (complexified) Euclidean group.  This particular 
coset model has not been  investigated in great detail, but Ghezelbash 
\cite{Ghezelbash} has argued that the resulting conformal field theory is 
a Liouville theory
\begin{align}
I_{\hbox{\scriptsize Liou}} = \frac{k}{8\pi}\int d^2x\sqrt{g}
  \left(\frac{1}{2}g^{ab}\partial_a\varphi\partial_b\varphi 
  +\frac{1}{2}\varphi R + \lambda e^\varphi\right)
\label{c3}
\end{align}
with a central charge $c=6k$.  This result is semiclassical, and
the setting is not quite right for loop quantum gravity, since the connection 
in \cite{Ghezelbash} is not self-dual.  But the result is at least suggestive.
Coset models based on $\mathrm{SL}(2,\mathbb{R})$ have been 
considered much more carefully, and they also, much less ambiguously, 
reduce to Liouville theory with the same central charge \cite{Forgacs,%
AlekShat,Bershadsky,Papa,Martinec,Balog,OR}.

Liouville theory is a well-known conformal field theory, which has been
widely studied \cite{Seiberg,Teschner} but is still not completely 
understood.  The quantum states come in two sets, commonly
called the ``normalizable'' and ``nonnormalizable'' or ``microscopic''
and ``macroscopic'' sectors.  The states in the normalizable sector have
conformal weights that are bounded below by finite values,
$$\Delta_{\hbox{\scriptsize\it min}} 
= {\bar\Delta}_{\hbox{\scriptsize\it min}} = \frac{c-1}{24} \ ,$$
so by (\ref{b2}), the effective central charges are the same as those 
of a free scalar field, 
$c_{\hbox{\scriptsize\it eff}} = {\bar c}_{\hbox{\scriptsize\it eff}} =1$.
The quantization of this sector is relatively well understood.  The 
nonnormalizable sector, on the other hand, has conformal weights
 that go down to zero, so the effective central charge is large for
large $k$, but the quantization of this sector is much more poorly 
understood (though see \cite{Chen} for one interesting attempt).  It is
suggestive that the normalizable sector corresponds classically to 
black-hole-like geometries, while the more mysterious nonnormalizable 
sector corresponds to point-particle-like geometries of the sort that 
arise in descriptions of spin network intersections with a black hole 
horizon \cite{Seiberg,Chen,Krasnov}.

\section{The three-dimensional gravity connection \label{grav}}

We may be able to learn more about the boundary action (\ref{a1}) by taking 
advantage of an accidental relationship with three-dimensional gravity.  In 
three-dimensional asymptotically anti-de Sitter space with a cosmological 
constant $\Lambda = -1/\ell^2$, the triad $e^a=e_\mu{}^adx^\mu$ and spin 
connection $\omega^a = \frac{1}{2}\epsilon^{abc}\omega_{\mu ab}dx^\mu$
can be combined as
\begin{align}
A^{(\pm)a} = \omega^a \pm \frac{\sqrt{-\sigma}}{\ell}e^a
\label{x1}
\end{align}
where $\sigma$ is the signature of the metric ($+1$ for Riemannian,
$-1$ for Lorentzian).  As noted by Achucarro and Townsend \cite{Achu}
and elaborated by Witten \cite{Wittenx}, the Einstein-Hilbert action 
then becomes a difference of Chern-Simons actions for $A^{(\pm)}$,
with coupling constants $k = \ell\sqrt{-\sigma}/4G_3$.  In particular, 
the boundary action (\ref{a1}) with gauge group $\mathrm{SL}(2,\mathbb{C})$ 
is equivalent to the action for three-dimensional Euclidean anti-de Sitter 
gravity.  For the self-dual (or ``holomorphic'') case, the  action includes 
an extra ``exotic'' term \cite{Wittenx}, but the field equations are unaffected.  
Similarly, the boundary action with gauge group $\mathrm{SL}(2,\mathbb{R})$ 
is equivalent to a chiral half of the action for three-dimensional Lorentzian 
anti-de Sitter gravity.   

Now, Brown and Henneaux showed long ago \cite{BrownHenneaux}
that the asymptotic symmetries of three-dimensional anti-de Sitter 
gravity are conformal symmetries, described by a Virasoro algebra 
with a central charge 
\begin{align}
c = {\bar c} = \frac{3\ell}{2G_3} = \frac{6k}{\sqrt{-\sigma}}   .
\label{x2}
\end{align}
While the microscopic degrees 
of freedom are not very well understood \cite{Carlipbh}, we have very 
good reasons to believe this result, especially since the Cardy formula 
then gives the correct entropy for the three-dimensional BTZ black hole.

To be sure, the physical settings for conformal symmetry differ for the
two theories.  In  three-dimensional gravity, the relevant symmetry group 
appears at asymptotically anti-de Sitter boundaries at infinity, while 
in loop quantum gravity we are interested in boundaries in the form 
of punctures or holes at the event horizon.   But the two geometries 
are related by conformal transformations, and the relevant conformal 
weights in three-dimensional gravity are determined by holonomies 
of the connection, which are invariant under such transformations.  
If we can better understand the conformal weights in loop quantum 
gravity, it should be possible to translate these into the setting 
of three-dimensional gravity.

An alternative connection to (2+1)-dimensional gravity
may also be useful.  In addition to describing three-dimensional 
anti-de Sitter gravity with a Riemannian signature metric, an 
$\mathrm{SL}(2,\mathbb{C})$ Chern-Simons action describes 
(2+1)-dimensional de Sitter gravity with Lorentzian signature.
The counting of states in this theory is not quite as well understood,
but there is very strong evidence that it can again be described in
terms of a two-dimensional conformal field theory.  In \cite{CarlipA},
it is shown that this equivalence yields the correct black hole entropy
for loop quantum gravity in a rather straightforward way.

\section{Entropy from the canonical Cardy formula}

Let us suppose for now that the horizon boundary term in loop 
quantum gravity can be related to a Liouville theory, as suggested
in section {\ref{SL2}, and that the effective central charge is given 
by the full central charge $c=6k$ (that is, that we include the 
nonnormalizable sector).   Alternatively, we may assume that
the connection to three-dimensional gravity described in section
\ref{grav} gives the correct central charge.  To determine the 
significance of this central charge in the (3+1)-dimensional setting, 
we need the coefficient $k$ of the Chern-Simons action (\ref{a1}).  
This is given in refs.\ \cite{ABCK,ABK,ISO}:
\begin{align}
k = \frac{iA_\Delta}{8\pi G} ,
\label{d1}
\end{align}
where $A_\Delta$ is the prescribed area of the horizon.  The
appearance of an imaginary coupling constant is  slightly peculiar,
but it is expected for a holomorphic Chern-Simons theory, where 
it can be traced to the factor of $\sqrt{-\sigma}$ in (\ref{x1}).  The 
factor of $i$ disappears in the central charge (\ref{x2}), and one can 
again find a relationship with Liouville theory with central charge 
$6k$ \cite{Harlow}.  It is interesting to note that this expression
gives one-half of the central charge obtained by looking at 
boundary terms in the generators of diffeomorphisms \cite{Ccov,CKerr}.

We also need the dimensionless temperature $T$.  The natural 
choice is the ``geometric temperature'' \cite{Pranzetti}, the 
temperature for which the horizon state is a KMS state,
\begin{align}
T = \frac{1}{2\pi\left(1-\frac{1}{k}\right)} \approx \frac{1}{2\pi}  .
\label{d2}
\end{align}
More simply, in the large $k$ limit for which our analysis might
be trusted, this is simply the local temperature seen by a
preferred quasilocal observer a proper distance $\ell$ from the 
horizon, scaled by $\ell$ to become dimensionless \cite{Froddenb}.

We can now insert these values into the canonical Cardy formula 
(\ref{b2}).  We obtain a contribution of
\begin{align}
S = \frac{\pi^2}{3}c_{\hbox{\scriptsize\it eff}}\,T
   = \frac{A_\Delta}{8G} ,
\label{d3}
\end{align}
one-half of the Bekenstein-Hawking entropy, from each
boundary of $\Delta$ upon which our conformal field theory
appears.

This result is somewhat less informative than one might hope.
It requires that the horizon $\Delta$ have a boundary---the WZNW 
model needs some two-manifold to live on---but the result depends 
on no particular characteristic of that boundary.  Rather, we learn 
that whatever the boundary, equilibrium at the temperature (\ref{d2}) 
requires that the conformal field theory be excited in such a way 
that the density of states is given by (\ref{d3}).  

In an eternal black hole, for instance, one possibility is to consider 
the bifurcation sphere to be an initial boundary of $\Delta$, with
a second boundary at the infinite future.  This is a somewhat
unusual picture, since the boundaries are spacelike, but 
Chern-Simons theory is a topological field theory, and may
not care about the distinction between spacelike and timelike
boundaries.  In this case, one gets a contribution (\ref{d3})
occurring twice, giving the full Bekenstein-Hawking entropy.
Another possibility, currently under investigation, is that the
relationship with three-dimensional gravity may permit a formal 
identification of the entropy (\ref{d3}) with that of one of
the two asymptotic regions of a three-dimensional black hole.%
\footnote{One might also look for a factor of two by considering a
nonchiral WZNW model, with contributions from left- and
right-moving modes.  But the self-dual/holomorphic connection
formalism leads much more naturally to a chiral model.}

\section{Counting states with the microcanonical Cardy formula \label{micro}}

While the canonical result for black hole entropy is interesting, it would
be nice to know more about the actual physical nature of the states.  To  
do so is more difficult, and may require one added assumption.  

In the usual approach to black hole entropy in loop quantum
gravity, horizon states are associated with punctures where
spin network edges cross the horizon.   Following \cite{GhoshPran},
let us enlarge these punctures to ``holes'' with finite, 
although arbitrarily small, boundaries.  As discussed in section
\ref{CSWZW}, the  boundary states then become the states
of a conformal field theory on a cylinder.   

As noted in section \ref{grav}, we do not really know the relevant
states from first principles.  We do know that spin network punctures
depend on holonomies, and in the three-dimensional gravity model
these holonomies determine the conformal weights.  From past work
involving analytic continuation of the partition function, the states 
that seem to be relevant to counting spin network edges \cite{Geiller,Achour} 
are classified by the unitary series of $\mathrm{SL}(2,\mathbb{C})$
representations, with  $j = \frac{1}{2}+is$.  These occur in the
relevant WZNW models (see section 4 of \cite{MalOog}, sections 
1--2 of \cite{MalOogb}, or \cite{Gaw}), and have conformal weights
\begin{align}
\Delta_j = -\frac{j(j-1)}{k-2} .
\label{e1}
\end{align}
Similarly, Liouville theory contains states with weights \cite{Chen,Gervais}
\begin{align}
\Delta_j = \frac{c-1}{24}-\frac{j(j-1)}{k-2}  .
\label{e1a}
\end{align}
(These can also be obtained from \cite{Seiberg} by the substitution
$j\rightarrow j-\frac{{\tilde k}+2}{2}$.)  The weights (\ref{e1}) and (\ref{e1a})
differ by a constant, but the Liouville values were obtained for the theory 
on the plane; as noted in section \ref{VA}, one must first subtract 
$c/24$ before using the Cardy formula.  

If we now insert these conformal weights into the microcanonical 
Cardy formula (\ref{b3}), and again assume a central charge $6k$,
we obtain a contribution of
\begin{align}
S \approx 2\pi \sum_{\hbox{\scriptsize punctures}}\!\!\!\sqrt{-j(j-1)} .
\label{e2}
\end{align}
(The square root is real for $j = \frac{1}{2}+is$.)  This is almost precisely
the standard formula for area in loop quantum gravity \cite{RovSmolin,AshLew}.
In fact, following the analytic continuation arguments of \cite{Geiller,Achour},
the right-hand side can be interpreted as exactly one quarter of the 
horizon area in loop quantum gravity, yielding the correct Bekenstein-Hawking
entropy.   A similar expression for the area can be obtained from the 
Alexandrov's Lorentz-covariant quantization \cite{Alexa,Alexb}, although 
there the area acquires an extra factor of $2$, becoming $16\pi G\sqrt{s^2+1/4}$
and making (\ref{e2}) one-half of the Bekenstein-Hawking entropy.
 
\section{A few inconclusive conclusions}

To conclude,
let us first see how the program sketched here contrasts with the standard
loop quantum gravity approach to black hole thermodynamics.  I believe
there are three essential differences: the choice of gauge group, the 
treatment of spin network intersections with the horizon, and the
role of combinatorics.

The first of these---the shift from $\mathrm{SU}(2)$ to
$\mathrm{SL}(2,\mathbb{C})$ or $\mathrm{SL}(2,\mathbb{R})$---is
suggested by a number of recent results in which the Barbero-Immirzi
parameter is analytically continued to its self-dual value $\gamma=i$.
This step is obviously problematic, since we do not yet really
understand the quantization of the theory when $\gamma$ is not
real, but perhaps the lesson is that we need to develop a better
understanding.   

The second---the treatment of spin network intersections with
the horizon as ``holes'' rather than ``punctures''---is less clear.   
The usual identification of punctures comes in several steps.  
First, isolated horizon boundary conditions require that at a fixed 
time $t$, the fields at the horizon satisfy
\begin{align}
F^{KL} \triangleq -\frac{\pi(1-\gamma^2)}{A_H}\Sigma^{KL}
\label{f1}
\end{align}
where $F^{KL}$ is the curvature of the self-dual connection $A^{KL}$,
$\Sigma^{KL} = \frac{1}{2}\epsilon^{KL}{}_{MN}e^M\wedge e^N$,
and the differential forms are pulled back to the horizon.  
When the right-hand side of (\ref{f1}) is evaluated on a bulk
spin network state, it gives a sum of delta functions, one at each
point where a spin network edge intersects the horizon, each multiplied 
by a generator $J^{KL}$ of the gauge group.  Eqn.\ (\ref{f1}) is
then a condition on the curvature $F$, which can be recognized 
as precisely the constraint equation for a Chern-Simons 
theory at time $t$ with a set of Wilson lines carrying the same
representations as the spin network edges.  The WZNW state is 
thus induced by Wilson lines rather than finite boundaries.%
\footnote{Note that the Wilson lines are \emph{not} the same 
as the spin network edges: the Wilson lines live inside the 
three-manifold $\Delta$, while the edges merely intersect 
$\Delta$.  One  might visualize a Wilson line in terms of 
spin foams, as a \emph{history} of an intersection of an
edge with the horizon.}

Observe, however, that $\Sigma^{KL}$ is not, strictly speaking, a
well-defined operator in loop quantum gravity: it is distributional,
and one should really deal only with integrals ${\hat\Sigma}_T$ over 
surfaces $T$ \cite{ABK,AshLew}.  Here, the only relevant surfaces
are portions of the horizon itself.  Classically, of course, the horizon
is a continuous surface, but quantum mechanically this is not at all
clear.  Since only integrals over ``plaquettes'' are well-defined in
the quantum theory, one might argue that (\ref{f1}) should hold 
only in integrated form.  

If this is the case, it is no longer obvious whether spin network 
intersections should be treated as punctures or as holes.
One way to pose this question, as described in section \ref{CSWZW}, 
is to ask whether any residual gauge transformations act on
loops surrounding an intersection.  It is not clear to me whether the
theory as currently formulated is capable of answering this question.

Finally, note that at least in section \ref{micro}, the entropy 
(\ref{e2}) is that of a \emph{fixed} set of punctures, and does
not include combinatorial factors from the different ways 
spin network punctures could give the same horizon area.  In
contrast, in the standard approach of \cite{ABCK}, the entropy
arises entirely from such combinatorics.  Indeed, the strange 
standard value of the Barbero-Immirzi parameter can be traced 
to the properties of this combinatorial problem.  

While the counting of states in a conformal field theory
is also a combinatorial problem, it is a rather different one,
essentially the determination of the number of partitions of a
large integer \cite{Carlipc}.  This is one of the few such problems
whose answer is a natural exponential (the Cardy formula),  leading
to a simple dependence on the Barbero-Immirzi parameter.
But if the ideas presented here are correct, one must understand 
why the usual combinatorial factor is either absent or subleading.
One possibility \cite{Alexc} is that different choices of punctures
correspond to genuinely different macroscopic states, since
the bulk geometries that can be consistently attached are also
different in the neighborhood of the punctures. This interpretation
gains support from the successes of canonical approaches that
treat the number of punctures as a new thermodynamic state 
variable \cite{Ghosh}.  In essence, this becomes question of how 
much coarse-graining is required to define the entropy of a
black hole. 

Even if these problems can be addressed, a good deal remains to be 
done.  It is not obvious which Chern-Simons theory is relevant: an 
$\mathrm{SL}(2,\mathbb{R})$ gauge group would simplify life, since 
many more results are known, but I do not know how to pick out a 
particular $\mathrm{SL}(2,\mathbb{R})$ factor from the full self-dual 
$\mathrm{SL}(2,\mathbb{C})$ action.  If the $\mathrm{SL}(2,\mathbb{C})$ 
Chern-Simons theory is indeed the correct one, many questions 
remain open, from the basic properties of a holomorphic Chern-Simons 
theory to the inadequately understood coset 
constructions.  If the end result is a Liouville theory, much is
still to be understood about the nonnormalizable sector, 
which must be included for the Cardy formula to give the correct 
entropy.  It has been argued in a different context that 
Liouville theory should be viewed as an effective description of   
more fundamental degrees of freedom \cite{Martinecb}; this issue 
remains unresolved.

Finally, it would be valuable to understand the relationship between
the approach described here and the more general attempts,
summarized in \cite{CEntropy}, to understand the entropy of
arbitrary black holes.  Both rely on two-dimensional conformal field 
theory, but at first sight the field theories are quite different: the
WZNW models and Liouville theory I have described here live on
the ``$\varphi$--$v$ cylinder,'' while the models considered in
\cite{CEntropy} live on the ``$r$--$v$ plane.''  A similar mismatch
occurs in the attempt to define a local temperature at the horizon
in loop quantum gravity \cite{Pranzetti}.  There, the simplicity
constraints (\ref{c0}) play a crucial role by connecting spatial
rotations to boosts, thus tying together symmetries in different
planes. Pranzetti has suggested that a similar mechanism may be
at work here \cite{Pranzettib}, but this, like much else in this 
paper, remains speculative.

The hints I have presented here thus raise as many questions as
they offer answers.  I believe, though, that some of these are 
at least the right questions.

\vspace{1.5ex}
\begin{flushleft}
\large\bf Acknowledgments
\end{flushleft}

I would like to thank Sergei Alexandrov, Abhay Ashtekar, Marc 
Geiller, Kirill Krasnov, and
Daniele Pranzetti for very helpful comments and conversations.
This work was supported in part by Department of Energy grant
DE-FG02-91ER40674.


\begin{thebibliography}{99}
\bibitem{Rovelli} C.\ Rovelli, Black hole entropy from loop quantum 
gravity, {Phys.\ Rev.\ Lett.} {77} (1996) 3288, arXiv:gr-qc/9603063.

\bibitem{ABCK} A.\ Ashtekar, J.~C.\ Baez, A.\ Corichi, and K.\ Krasnov,
Quantum geometry and black hole entropy, {Phys.\ Rev.\ Lett.}
{80} (1998) 904, arXiv:gr-qc/9710007.

\bibitem{ABK} A.\ Ashtekar, J.~C.\ Baez, and K.\ Krasnov, 
Quantum geometry of isolated horizons and black hole entropy,
{Adv.\ Theor.\ Math.\ Phys.} {4} (2000) 1, 
arXiv:gr-qc/0005126.

\bibitem{Domagala} M.\ Domagala and J.\ Lewandowski, Black hole 
entropy from quantum geometry, {Class.\ Quant.\  Grav.}
{21} (2004) 5233, arXiv:gr-qc/0407051.

\bibitem{Meissner} K.~A.\ Meissner, Black hole entropy in loop 
quantum gravity, {Class.\ Quant.\  Grav.} {21}
(2004)  5245, arXiv:gr-qc/0407052. 

\bibitem{Ghosh}  A.\ Ghosh and A.\ Perez, Black hole entropy and
isolated horizons thermodynamics, {Phys.\ Rev.\ Lett.} {27}
(2011) 241301, arXiv:1107.1320.

\bibitem{Perez} A.\ Ghosh, K.\ Noui, and A.\ Perez, Statistics, 
holography, and black hole entropy in loop quantum gravity,
arXiv:1309.4563.

\bibitem{GhoshPran} A.\ Ghosh and D.\ Pranzetti, CFT/gravity 
correspondence on the isolated horizon, arXiv:1405.7056.

\bibitem{Geiller} E.\ Frodden, M.\ Geiller, K.\ Noui, and A.\ Perez,
Black hole entropy from complex Ashtekar variables, {Europhys.\
Lett.} {107} (2014) 10005, arXiv:1212.4060.

\bibitem{Achour} J.~B.\ Achour, A.\ Mouchet, and K.\ Noui,
Analytic continuation of black hole entropy in loop quantum 
gravity, arXiv:1406.6021. 

\bibitem{Bodendorfer} N.\ Bodendorfer and Y.\ Neiman,
Imaginary action, spinfoam asymptotics and the `transplanckian' 
regime of loop quantum gravity, {Class.\ Quantum Grav.} {30} 
(2013) 195018, arXiv:1303.4752.

\bibitem{Ashtekar} A.\ Ashtekar, New variables for classical and 
quantum gravity, {Phys.\ Rev. Lett.} {57}  (1986), 2244.

\bibitem{Samuel} J.\ Samuel, Is Barbero's Hamiltonian formulation 
a gauge theory of Lorentzian gravity?, {Class.\ Quant.\  Grav.}
{17} (2000) L141, arXiv:gr-qc/0005095.

\bibitem{Alexa} S.\ Alexandrov, On choice of connection in loop 
quantum gravity, {Phys.\ Rev.\ D} {65} (2002) 024011,
arXiv:gr-qc/0107071.

\bibitem{Bianchi} E.\ Bianchi, Entropy of non-extremal black 
holes from loop gravity, arXiv:1204.5122.

\bibitem{Smolin0} L.\ Smolin, Linking topological quantum field 
theory and nonperturbative quantum gravity, {J.\ Math.\ Phys.}
{36} (1995) 6417, arXiv:gr-qc/9505028.

\bibitem{Smolinb} L.\ Smolin, A holographic formulation of 
quantum general relativity, {Phys.\ Rev.} {D61} (2000) 084007,
arXiv:hep-th/9808191.

\bibitem{Haggard} H.~M.\ Haggard, M.\ Han, W.\  Kami{\'n}ski,
and A.\ Riello, SL(2,C) Chern-Simons theory, a non-planar graph 
operator, and 4D loop quantum gravity with a cosmological constant: 
semiclassical geometry, arXiv:1412.7546.

\bibitem{ISO} A.\ Ashtekar,  A.\ Corichi, and K.\ Krasnov,
Isolated horizons: the classical phase space, {Adv.\ Theor.\ Math.\
Phys.} {3} (1999) 419, arXiv:gr-qc/9905089.

\bibitem{Krasnov0} K.\ Krasnov, On Quantum statistical mechanics 
of Schwarzschild black hole, {Gen.\ Rel.\ Grav.} {30} (1998) 53,
arXiv:gr-qc/9605047.

\bibitem{Wittend} E.\ Witten, Analytic continuation of Chern-Simons
theory, in \emph{Chern-Simons gauge theory: 20 years after},
edited by J.~E.\ Andersen, H.~U.\ Boden, A.\ Hahn, and B.\ Himpel 
(Am.\ Math.\ Soc., Providence, 2011), p.\ 347, arXiv:1001.2933.

\bibitem{Wittena} E.\ Witten, Quantum field theory and the
Jones polynomial, Commun.\ Math.\ Phys.\ 121 (1989) 351.

\bibitem{Ogura} W.\ Ogura, Path integral quantization of 
Chern-Simons gauge theory, Phys.\ Lett.\ B229 (1989) 61.

\bibitem{Carlipi} S.\ Carlip, Inducing Liouville theory from 
topologically massive gravity, Nucl.\ Phys.\ B362 (1991) 111.

\bibitem{Carlipa} S.\ Carlip, Statistical mechanics and 
black hole thermodynamics, {Nucl.\ Phys.\ Proc.\ Suppl.}
{57} (1997) 8, arXiv:gr-qc/9702017.

\bibitem{Wittenb} E.\ Witten, On holomorphic factorization of 
WZW and coset models, Commun.\ Math.\ Phys.\ 144 (1992) 
189.

\bibitem{Wittenf} E.\ Witten, Non-abelian bosonization in two 
dimensions, {Commun.\ Math.\ Phys.} {92} (1984) 455.

\bibitem{EMSS} S.\ Elitzur, G.\ W.\ Moore, A.\ Schwimmer, and N.\ Seiberg,
Remarks on the canonical quantization of the Chern-Simons-Witten 
theory, Nucl.\ Phys.\ B326 (1989) 108.

\bibitem{FMS} P.\ Di Francesco, P.\ Mathieu, and D.\ S{\'e}n{\'e}chal,
\emph{Conformal field theory} (Springer, New York, 1997).

\bibitem{Cardy} J.~A.\ Cardy, Operator content of two-dimensional 
conformally invariant theories, {Nucl.\ Phys.\ B} {270}
(1986) 186.

\bibitem{Cardy2}  H.~W.~J.\ Bl{\"o}te, J.~A.\ Cardy, and M.~P.\ 
Nightingale, Conformal invariance, the central charge, and 
universal finite size amplitudes at criticality, {Phys.\ Rev.\ 
Lett.} {56} (1986) 742.

\bibitem{Carlipc} S.\ Carlip, What we don't know about BTZ black 
hole entropy, {Class.\ Quant.\  Grav.} {15} (1998),
3609, arXiv:hep-th/9806026, section 2.

\bibitem{BoussoM} R.\ Bousso, A.\ Maloney, and A.\ Strominger,
Conformal vacua and entropy in de Sitter space, {Phys.\ Rev.} {D65}
(2002) 104039, arXiv:hep-th/0112218, section 8.

\bibitem{ENPP} J.\ Engle, K.\ Noui, A.\ Perez, and D.\ Pranzetti,
The SU(2) Black Hole entropy revisited, {JHEP} {1105} (2011) 016,
arXiv:1103.2723.

\bibitem{EPRL} J.\ Engle, E.\ Livine, R.\ Pereira, and C.\ Rovelli,
LQG vertex with finite Immirzi parameter, {Nucl.\ Phys.\ B}
{799} (2008) 136, arXiv:0711.0146.

\bibitem{Kaul} R.~K.\ Kaul and P.\ Majumdar, Schwarzschild 
horizon dynamics and SU(2) Chern-Simons theory, {Phys.\ Rev.} 
{D83} (2011) 024038, arXiv:1004.5487.

\bibitem{Wang} J.\ Wang and Y.\ Ma, BF theory explanation of 
the entropy for nonrotating isolated horizons, {Phys.\ Rev.} 
{D89} (2014) 084065, arXiv:1401.2967. 

\bibitem{Wittene} E.\ Witten, Quantization of Chern-Simons gauge 
theory with complex gauge group, {Commun.\ Math.\ Phys.} {137}
(1991) 29.

\bibitem{MalOog} J.~M.\ Maldacena and H.\ Ooguri, Strings in AdS(3) 
and SL(2,R) WZW model 1: the spectrum, {J.\ Math.\ Phys.} {42} (2001)
2929, arXiv:hep-th/0001053.

\bibitem{MalOogb}  J.~M.\ Maldacena and H.\ Ooguri, Strings in AdS(3) 
and SL(2,R) WZW model 3: correlation functions, {Phys.\ Rev.\ D} {65} 
(2002) 106006, arXiv:hep-th/0111180.

\bibitem{Gawedzki} K.\ Gawedzki and A.\ Kupiainen, Coset construction 
from functional integrals, {Nucl.\ Phys.\ B} {320} (1989) 625.

\bibitem{IH} A.\ Ashtekar, S.\ Fairhurst, and B.\ Krishnan,
Isolated horizons: Hamiltonian evolution and the first law,
{Phys.\ Rev.\ D} {62} (2000) 104025, arXiv:gr-qc/0005083.

\bibitem{Ghezelbash} A.~M.\ Ghezelbash, Gauging of Lorentz 
group WZW model by its null subgroup, {Mod.\ Phys.\ Lett.\ A}
{11} (1996) 1765, arXiv:hep-th/9607084.

\bibitem{Forgacs} P.\ Forgacs, A.\ Wipf, J.\ Balog, L.\ Feher, and
 L.\ O'Raifeartaigh, Liouville and Toda theories as conformally 
reduced WZNW theories, Phys.\ Lett.\ B227 (1989) 214.

\bibitem{AlekShat} A. Alekseev and S. Shatashvili, Path integral 
quantization of the coadjoint orbits of the Virasoro group and 
2D gravity, Nucl.\ Phys.\ B323 (1989) 719.

\bibitem{Bershadsky}  M.\ Bershadsky and H.\ Ooguri, Hidden
SL(n) symmetry in conformal field theories, Commun.\ Math.\ Phys.\
126 (1989) 49.

\bibitem{Papa} G.\ Papadopoulos and B.\ Spence, A covariant 
canonical description of Liouville field theory, Phys.\ Lett.\ B308 
(1993) 253, arXiv:hep-th/9303076.

\bibitem{Martinec} E.~J.\ Martinec and S.~L.\ Shatashvili, Black hole 
physics and Liouville theory, {Nucl.\ Phys.\ B} {368} (1992) 338.

\bibitem{Balog} J.\ Balog, L.\ Feher, and L.\ Palla, Coadjoint orbits 
of the Virasoro algebra and the global Liouville equation, Int.\ J.\ 
Mod.\ Phys.\ A13 (1998) 315, arXiv:hep-th/9703045.

\bibitem{OR} L.\ O'Raifeartaigh and V.\ V.\ Sreedhar, Path integral 
formulation of the conformal Wess-Zumino-Witten $\rightarrow$ 
Liouville reduction, Phys.\ Lett.\ B425 (1998) 291, 
arXiv:hep-th/9802077.

\bibitem{Seiberg} N.\ Seiberg, Notes on quantum Liouville theory and 
quantum gravity, {Prog.\ Theor.\ Phys.\ Suppl.} {102} (1990) 319.

\bibitem{Teschner} J.\ Teschner, Liouville theory revisited, {Class.\ 
Quant.\  Grav.} {18} (2001) R153, arXiv:hep-th/0104158.

\bibitem{Chen} Y.\ Chen, Quantum  Liouville theory and BTZ black 
hole entropy, {Class.\ Quantum  Grav.} {21} (2004) 1153,
arXiv:hep-th/0310234.

\bibitem{Krasnov} K.\ Krasnov, 3-D gravity, point particles and 
Liouville theory, {Class.\ Quantum Grav.} {18} (2001) 1291,
arXiv: hep-th/0008253.

\bibitem{Achu} A.\ Achucarro and P.~K.\ Townsend, A Chern-Simons 
action for three-dimensional anti-de Sitter supergravity theories,
{Phys.\ Lett.\ B} {180} (1986) 89.

\bibitem{Wittenx} E.\ Witten, 2+1 dimensional gravity as an exactly
soluble system, {Nucl.\ Phys.\ B} 311 (1998) 46.

\bibitem{BrownHenneaux} J.~D.\ Brown and M.\ Henneaux, Central 
charges in the canonical realization of asymptotic symmetries: an 
example from three-dimensional gravity, {Commun.\ Math.\ Phys.} 
{104} (1986) 207.

\bibitem{Carlipbh} S.\ Carlip, Conformal field theory, (2+1)-dimensional 
gravity, and the BTZ black hole, {Class.\ Quantum Grav.} {22} (2005)
R85, arXiv:gr-qc/0503022.

\bibitem{Harlow} D.\ Harlow, J.\ Maltz, and E.\ Witten, Analytic 
continuation of Liouville theory, {JHEP} {1112} (2011) 071,
arXiv:1108.4417.

\bibitem{CarlipA} S.\ Carlip, Four-dimensional entropy from 
three-dimensional gravity, arXiv:1503.02981.

\bibitem{Ccov} S.\ Carlip, Entropy from conformal field theory at 
Killing horizons, {Class.\ Quantum Grav.} {16} (1999) 3327,
arXiv:gr-qc/9906126.

\bibitem{CKerr} S.\ Carlip, Extremal and nonextremal Kerr/CFT 
correspondences, {JHEP} {1104} (2011) 076, arXiv:1101.5136.

\bibitem{Pranzetti} D.\ Pranzetti, Geometric temperature and entropy 
of quantum isolated horizons, {Phys.\ Rev.\ D} {89} (2014) 104046,
arXiv:1305.6714.

\bibitem{Froddenb} E.\ Frodden, A.\ Ghosh, and A.\ Perez, Quasilocal 
first law for black hole thermodynamics, {Phys.\ Rev.\ D} {87} (2013) 
121503, arXiv:1110.4055.

\bibitem{Gaw} K.\ Gawedzki, Non-Compact WZW conformal field 
theories, in \emph{Cargese 1991, Proceedings, New symmetry 
principles in quantum field theory}, edited by J.\ Frolich et al.\
(Plenum Press, New York, 1992), p.\ 247, arXiv:hep-th/9110076.

\bibitem{Gervais} J.-L.\ Gervais, The quantum group structure of 2D 
gravity and minimal models I, {Commun.\ Math.\ Phys.} {130} (1990)

\bibitem{RovSmolin} C.\ Rovelli and L.\ Smolin, Discreteness of area 
and volume in quantum gravity, {Nucl.\ Phys.} {B442} (1995) 593,
arXiv:gr-qc/9411005.

\bibitem{AshLew} A.\ Ashtekar and J.\ Lewandowski, Quantum theory 
of geometry. 1: Area operators, {Class.\ Quantum Grav.} {14} (1997) 
A55, arXiv:gr-qc/9602046.

\bibitem{Alexb}  S.\ Alexandrov and D.\ Vassilevich, Area spectrum 
in Lorentz covariant loop gravity, {Phys.\ Rev. D} {64} (2001) 
044023, arXiv:gr-qc/0103105.

\bibitem{Alexc} S.\ Alexandrov, On the counting of black hole states 
in loop quantum gravity, arXiv:gr-qc/0408033.

\bibitem{Martinecb} E.\ Martinec, Conformal field theory, 
geometry, and entropy, arXiv:hep-th/9809021.

\bibitem{CEntropy} S.\ Carlip, Effective conformal descriptions of 
black hole entropy, {Entropy} 13 (2011) 1355, arXiv:1107.2678.

\bibitem{Pranzettib}  D.\ Pranzetti, personal communication.


\end{thebibliography}
\end{document}